\documentclass[10pt]{article}
\title{Asset Price Dynamics in a Financial Market with Heterogeneous Trading Strategies and Time Delays}
\author{Alessandro Sansone$^{1}$\footnote{Corresponding author. $E-mail$ $address$:
alessandro.sansone@fastwebnet.it}, Giuseppe Garofalo$^2$,
\\  \normalsize $^1$ Department of Economic Sciences, \normalsize University of Rome "La Sapienza", Italy\\
\normalsize School of Finance and Economics, \normalsize University
of Technology, Sydney, NSW, Australia \\ \normalsize $^2$Department
of Managerial, Technological and Quantitative Studies\\
\normalsize University of Tuscia, Viterbo, Italy and
\\ \normalsize Department of Public Economics, \normalsize
University of Rome "La Sapienza", Italy}

\usepackage{graphicx}

\begin{document}
\maketitle
\begin{abstract}In this paper we present a continuous time dynamical model of
  heterogeneous agents interacting in a financial market where transactions are
  cleared by a market maker. The market is composed of fundamentalist, trend following
  and contrarian agents who process information from the market with different time delays.
  Each class of investor is characterized by path dependent risk aversion. We also allow for
  the possibility of evolutionary switching between trend following and contrarian strategies.
  We find that the system shows periodic, quasi-periodic and chaotic dynamics as well as
  synchronization between technical traders. Furthermore, the model is able to generate
   time series of returns that exhibit statistical properties similar to those of the S$\&$P500
   index, which is characterized by excess kurtosis, volatility clustering and long memory.
\end{abstract}


\section{\label{sec:introduction}Introduction}
In recent years there has been a growing disaffection with the
standard paradigm of efficient markets and rational expectations. In
an efficient market, asset prices are the outcome of the trading of
rational agents, who forecast the expected price by exploiting all
the information available and know that other traders are rational.
This implies that prices must equal the fundamental prices and
therefore changes in prices are only caused by changes in the
fundamental value. In real markets, however, traders have different
information on traded assets and process information differently,
therefore the assumption of homogeneous rational traders may not be
appropriate. The efficient market hypothesis motivates the use of
random walk increments in financial time series modeling: if news
about fundamentals are normally distributed, the returns on an asset
will be normal as well. However the random walk assumption does not
allow the replication of some stylized facts of real financial
markets, such as volatility clustering, excess kurtosis,
autocorrelation in square and absolute returns, bubbles and crashes.
Recently a large number of models that take into account
heterogeneity in financial markets has been proposed. contributions
to this literature include \cite{beja,BH98,chiarella,CH,Wa,Wb}.
\cite{TDG} analyze a market composed of a continuum of
fundamentalists who show delays in information processing. These
models allow for the formation of speculative bubbles, which may be
triggered by news about fundamentals and reinforced by technical
trading. Because of the presence of nonlinearities according to
which different investors interact with one another, these models
are capable of generating stable equilibria, periodic,
quasi-periodic dynamics and strange attractors. This paper builds on
the model of \cite{TDG}, which is inspired by the models of
thermodynamics of \cite{hoover}, \cite{nosè84a}, \cite{nosè84b} and
analyzes a financial market in which there are only fundamental
investors who trade according to the mispricing of the asset with
delays which are uniformly distributed from initial to current time.
We generalize \cite{TDG} by introducing a continuum of technical
traders who behave as either trend followers or contrarians and a
switching rule between these technical trading rules. We will
analyze how the interaction of different types of investors with
path dependent risk aversion determines the dynamics and the
statistical properties of the system as key parameters are changed.

\section{\label{sec:Themodel}The model} Let us consider a security continuously traded at price
$P(t)$ Assume that this security is in fixed supply, so that the
price is only driven by excess demand. Let us assume that the excess
demand $D(t)$ is a function of the current price and the fundamental
value $F(t)$. A market maker takes a long position whenever the
excess demand is negative and a short position whenever the demand
excess is positive so as to clear the market. The market maker
adjusts the price in the direction of the excess demand with speed
equal to $\lambda^M$. The instantaneous rate of return is:
\begin{equation}\label{return}
r(t)\equiv \frac {\dot{P}(t)}{P(t)}=\lambda^M
D(P(t),F(t));\,\,\,\lambda^M>0.
\end{equation}
\noindent the fundamental value is assumed to grow at a constant
rate g, therefore:
\begin{equation}\label{fundamental}
\frac {\dot{F}(t)}{F(t)}=g.
\end{equation}
\noindent The market is composed of an infinite number of investors,
who choose among three different investment strategies. Let us
assume that a fraction $\alpha$ of investors follows a
fundamentalist strategy and a fraction $(1-\alpha)$ follows a
technical analysis strategy. The fraction of technical analysts is
in turn composed of a fraction $\beta$ of trend followers and a
fraction $(1-\beta)$ of contrarians. Let $D^F(t)$, $D^{TF}(t)$ and
$D^C(t)$ be respectively the demands of fundamentalists, trend
followers and contrarians rescaled by the proportions of agents who
trades according to a given strategy. The excess demand for the
security is thus given:
\begin{equation}\label{ED}
D(t)=\alpha D ^F(t)+(1-\alpha )[\beta D^{TF}(t)+(1-\beta ) D^
C(t)]\,;\,\,\alpha, \,\beta\,\epsilon\,[0,1].
\end{equation}
\noindent Each trader operate with a delay equal to $\tau$, that is,
the demand of a particular trader at time $t$ depends on her
decision variable at time $t-\tau$. Time delays are uniformly
distributed in the interval $[0,t]$. Fundamentalists react to
differences between price and fundamental value. The demand of
fundamentalists operating with delay $\tau$ is:
\begin{equation}\label{DFTAU}
D^{F\tau}(t)=\lambda^{F\tau}\log \biggl [\frac
{F(t-\tau)}{P(t-\tau)}\biggl ]\,;\,\lambda^{F\tau}>0
\end{equation}
\noindent where $\lambda^{F\tau}$ is a parameter that measures the
speed of reaction of fundamental traders; we will assume that
$\lambda^{F\tau}=\lambda^F$ throughout the paper. This demand
function implies that the fundamentalists believe that the price
tends to the fundamental value in the long run and reacts to the
percentage mispricing of the asset in symmetric way with respect to
underpricing and overpricing. If time delays are uniformly
distributed, the market demand of fundamentalists is given by:
\begin{equation}\label{DF}
D^F(t)=\lambda^F \int_0^t \log \biggl [\frac
{F(t-\tau)}{P(t-\tau)}\biggl ]\,d\tau \,;\,\lambda^F>0
\end{equation}
\noindent time differential yields:
\begin{equation}\label{DFD}
\dot {D}^F(t)=\lambda^F \log \biggl [\frac {F(t)}{P(t)}\biggl
]\,;\,\lambda^F>0.
\end{equation}
\noindent Following \cite{TDG}, let us modify equation (\ref{DFD})
by introducing the variable $\varsigma$ and adding a term
$-\varsigma^F(t)D^F(t)$ to the RHS:\footnote{\cite{TDG} introduce
the variable $\xi$, which is a liner transformation of $D^F(t)$, and
utilize it instead of $D^F(t)$ in the simulations. We will continue
to utilize $D^F(T)$ without any loss of generality.}
\begin{equation}\label{DFDR}
\dot {D}^F(t)=\lambda^F \log \biggl [\frac {F(t)}{P(t)}\biggl
]-\varsigma(t)^F D^F(t)\,;\,\lambda^F>0.
\end{equation}
\noindent According to the sign of $\varsigma$, if there is an
excess demand, the term either drives it towards zero (if
$\varsigma^F(t)$ is positive) or foster it (if $\varsigma^F(t)$ is
negative). The variable $\varsigma^F(t)$ may be interpreted as in
indicator of the risk that traders bear and their risk aversion (if
$\varsigma^F(t)$ is negative traders become risk-seekers). The
dynamics for $\varsigma^F(t)$ are given by:
\begin{equation}\label{DRF}
\dot {\varsigma(t)}^F(t)=\delta^ F[D^F(t)^2-V^F]\,;\,\delta^F>0
\end{equation}
\noindent where $V^F$ is a factor controlling the variance.
Throughout the paper we will assume that $V^F$ is given. The
rationale of (\ref{DRF}) is that the larger an open position on the
asset, the more risk averse the agents become. Let us consider now
the behavior of technical traders. As for the fundamentalists, their
time delays are uniformly distributed in the interval $[0,t]$. A
trader operating with delay $\tau$ utilizes the percentage return
that occurred at time $t-\tau$ in a linear prediction rule in order
to form an expectation of future returns. Let $D^{TF\tau}$ and
$D^{C\tau}$ be respectively the demands of trend followers and
contrarians operating with delay $\tau$. Without taking risk
attitudes into account, technical demands are given by:
\begin{equation}\label{DTTAU}
D^{i\tau}(t)=\lambda^{i\tau}\log [r(t-\tau
)]\,;\,i=TF,C\,;\,\lambda^{TF\tau}>0\,;\,\lambda^{C\tau}<0.
\end{equation}
\noindent Throughout the paper we will assume that
$D^{TF\tau}=D^{TF}$ and $D^{C\tau}=D^C$. By integrating
(\ref{DTTAU}) with respect to $\tau$, time differentiating and
adding respectively the terms $-\varsigma^{TF}(t)D^{TF}(t)$ and
$-\varsigma^C(t)D^C(t)$ in order to take into account the risk and
risk attitudes of technical traders, we get:
\begin{equation}\label{DT}
\dot D^i(t)=\lambda^i\log
[r(t)]-\varsigma^i(t)D^C(t)\,;\,i=TF,C\,;\,\lambda^{TF\tau}>0\,;\,\lambda^{C\tau}<0
\end{equation}
\noindent the dynamics for $-\varsigma^{TF}(t)$ and
$-\varsigma^C(t)$ have the same functional form as
$-\varsigma^F(t)$:
\begin{equation}\label{DRT}
\dot
{\varsigma}^i(t)=\delta^i[D^i(t)^2-V^i]\,;\,\delta^i>0\,;\,i=TF,C
\end{equation}
\noindent We will now consider the fraction $\alpha$ as given,
whereas the fraction of trend followers $\beta$ may be path
dependent. In fact, $\beta$ is considered as an endogenous variable
because both trend followers and contrarians follow technical
trading strategies and therefore may be likely to switch them if one
is more profitable than the other. We assume that the more
profitable is a strategy, the more investors will choose that
strategy. The difference in the absolute return at time $t$ between
the two strategies is given by $\dot
P(t)[D^{TF}(t)-D^C(t)]$.\footnote{The use of absolute returns as a
measure of evolutionary fitness stems from the absence of wealth in
the model, therefore it is not possible to define the percentage
return of a strategy.} Moreover, $\beta$ must be bounded in the
interval $[0,1]$ and we assume that it tends to move towards 0.5 if
both strategies lead to equal profits. These assumption hold if we
assume that dynamics for $\beta$ is the following:
\begin{equation}\label{S}
\dot \beta (t)=\cot[\pi\beta (t)]+z\dot P(t)[
D^{TF}(t)-D^C(t)]\,;\,\,z\geq0
\end{equation}
\noindent where the first term keeps the fraction of trend followers
bounded in the interval $[0,1]$ and $z$ is a parameter that measure
the speed of switching between the technical strategies. If z=0 or
if the proportion of trend followers and contrarians is taken as a
constant, then the system may be made stationary by defining the
variable $M(t)\equiv F(t)/P(t)$, whose time derivative is:
\begin{equation}\label{m}
\dot M(t)=g-\lambda^M\bigl[\alpha D ^F(t)+(1-\alpha )[\beta
D^{TF}(t)+(1-\beta ) D^ C(t)]\bigl].
\end{equation}

\section{\label{sec:Statprop}Statistical properties}

\begin{figure}
\ \

\includegraphics[width=11cm,height=6.83cm]{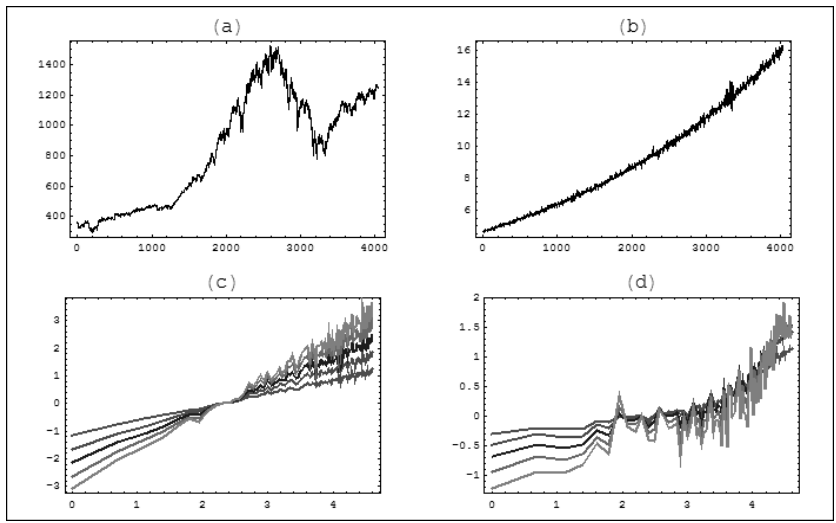}
\caption{Time series of S$\&$P500 (a), model-generated prices (b),
plot of $\log E[\mid x(t,\Delta t) \mid ^q]$ against $\log [\Delta
t]$ for S$\&$P500 (c) and simulations (d) respectively for
$q+1=1,1.5,2,2.5,3$ top down at the left side. } \label{F1}
\end{figure}

In this section, we analyze the statistical properties of the
simulated time series, which have been generated by integrating the
system up to time 9035 and recording the price at integer times
starting from $t=5000$ in order to allow the system to get
sufficiently close to the asymptotic dynamics and to have time
series as long as the daily time series of the S$\&$P500 index
between 1 January 1990 and 31 December 2005. The system has been
integrated by utilizing Mathematica 5.1. No stochastic elements are
added, therefore the features of system-generated time series are
endogenous and originate from the nonlinear structure of the
systems. The model displays statistical properties similar to those
of the index S$\&$P500 using various parameter values. In
\emph{Table (1)} there are reported the statistics of the daily
returns on the S$\&$P500 and on the time series generated by the
system with parameters $\lambda ^M=60$, $\lambda ^F=95/15$, $\lambda
^{TF} =0.25$, $\lambda ^C=-0.22$, $\alpha =0.6$, $\delta ^F=\delta
^{TF}=\delta ^C=240000$, $V^F=V^{TF}=V^C=54000$, $g=0.000308$, $z=0$
and initial values $P=1.1$, $F=1$ $D^F=\lambda ^F \log [G(0)/P(0)]$,
$D^{TF}=D^C=0$, $\varsigma ^F=\varsigma ^{TF}=\varsigma ^C=1$,
$\beta=0.5$. We have also reported the value of the largest Lyapunov
exponent. The growth rate of the fundamental, $g$, is equal to the
mean growth rate of S$\&$P500, which in turn has been calculated as
the rate that in a continuously compounded capitalization regime
implies the same return on the index on the overall period. Since
the price moves around the fundamental, the mean of the simulated
time series match that of the S$\&$P500. The other parameter values
have been chosen so as to give rise to statistics similar to those
of the S$\&$P500 index. As pointed out by \cite{TDG}, kurtosis and
volatility clustering are due to the delayed reaction of investors
that determines price overshooting. In a multi-agent modeling, such
a process is fostered by the interaction among investors who are
heterogeneous not only as concerns the time that they need to
process information from the market, but also the strategies that
they use to predict future prices. Time series are also
characterized by long memory and nonlinear structure, which in turn
imply that volatility clustering occurs at different time scales.
Such characteristics are typical of multifractal process. According
to \cite{MFC}, a multifractal process is a continuous time process
with stationary increments which satisfy:
\begin{equation}\label{MF}
E[\mid x(t,\Delta)\mid^q]=c(q)(\Delta t)^{\tau (q)+1}\,;\,
x(t,\Delta t)=x(t+\Delta t)-x(t)\,\,\,;\,\,\,0\le t \le T
\end{equation}
\noindent under existence conditions given in \cite{MFC}. Assuming
that $x(t)=\log P(t)-\log P(0)$ \emph{Table (2)} reports the $R^2$
of a regression of $\log E[\mid P(t,\Delta)\mid^q]$ against $\log
\Delta (t)$ with $q+1=1,1.5,2,2.5,3$. $P$ is the daily closure of
S$\&$P500 and the model-generated time series. \emph{Figure
(\ref{F1})} reports the time series and the $\log$-$\log$ plot after
normalizing by subtracting $\log E\mid x(t,\ log[10])\mid^q]$. Time
intervals range from 1 to 100 days. There is no apparent crossover
up to a scale of 100 days in the S$\&$P500 and the linear fit is
very good, in accord with the behavior of a multifractal process.
Crossover occurs in the simulations for values of $t$ between $e^3$
and $e^4$ and the fluctuations are more erratic than those of
S$\&$P500. Such a behavior underlines the capability of the model to
generate dynamics typical of a multifractal process, however the
dynamics for the fundamental implies that price is mean-reverting
around an exponenential trend, which in turn implies that crossover
occurs for smaller time intervals than those of real time series.
The introduction of stochastic noises or a feedback between
fundamental and price determines more a realistic long-run behavior
and scaling properties, as we will show for the latter case in
\emph{Section 4.4}.

\section{\label{sec:analysis}Sensitivity analysis}
\begin{figure}
\ \

\includegraphics[width=11cm,height=8.8cm]{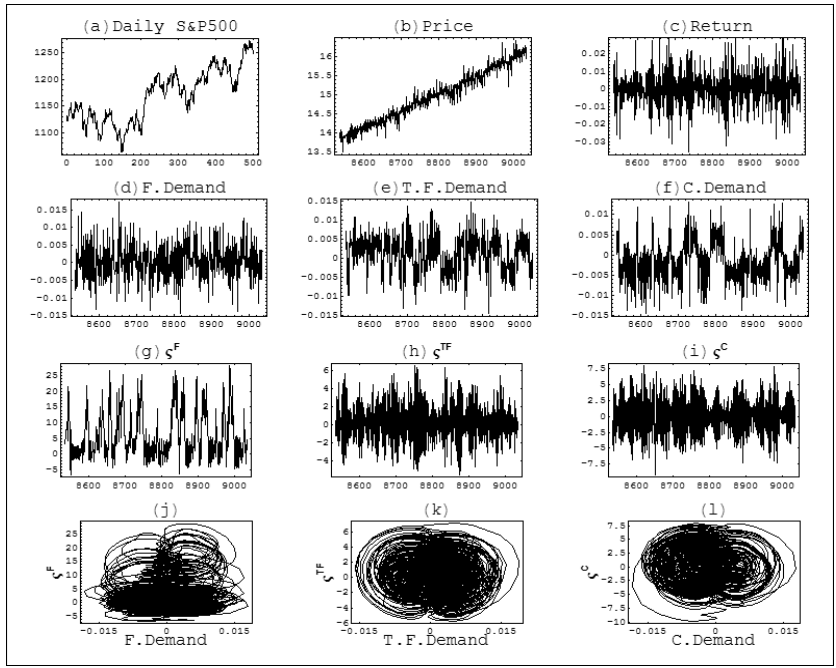}
\caption{Time series of S$\&$P500 (a), price (b), returns(c), demand
of fundamentalists (d), trend followers (e), contrarians (f), risk
aversion of fundamentalists (g), trend followers (h), contrarians
(i), projection of the phase space on planes $[D^F, \varsigma ^F]$
(j), $[D^{TF}, \varsigma ^{TF}]$ (k), $[D^C, \varsigma ^C]$ (l).}
\label{F2}
\end{figure}

\begin{figure}

\includegraphics[width=11cm,height=6.9cm]{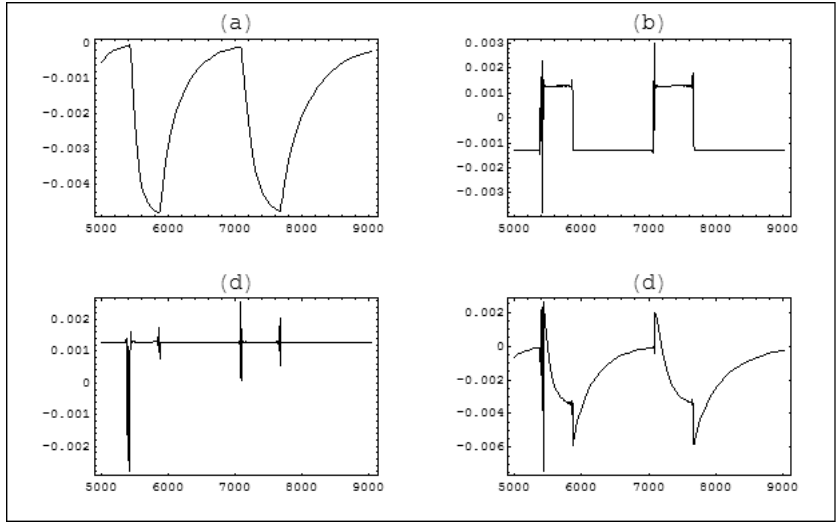}
\caption{Total demand of fundamentalists (a), trend followers (b),
contrarians (c) and market excess demand (d) when $\alpha =0.2$.}
\label{F3}
\end{figure}

\begin{figure}
\ \

\includegraphics[width=11cm,height=3.3cm]{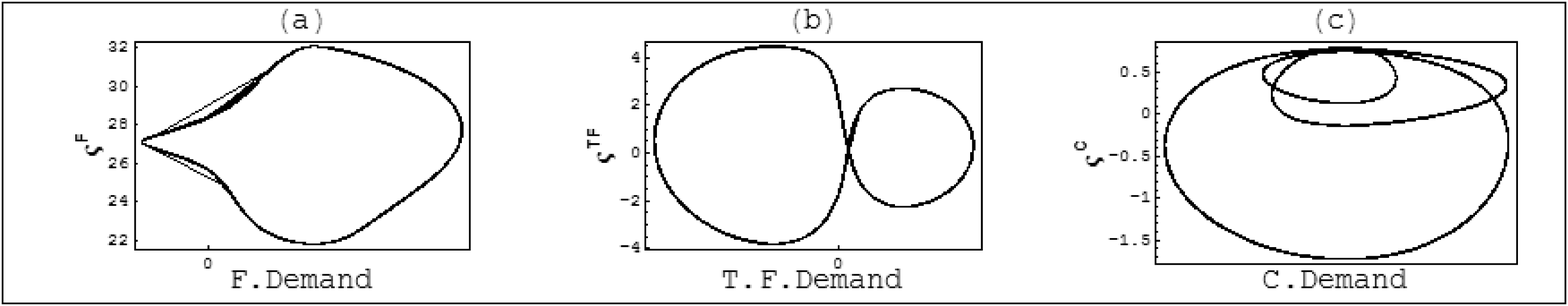}
\caption{projection of the phase space on planes $[D^F, \varsigma
^F]$ (j), $[D^{TF}, \varsigma ^{TF}]$ (k), $[D^C, \varsigma ^C]$ (l)
when $\lambda^ F =190$.} \label{F4}
\end{figure}

\begin{figure}
\ \
\includegraphics[width=11cm,height=3.3cm]{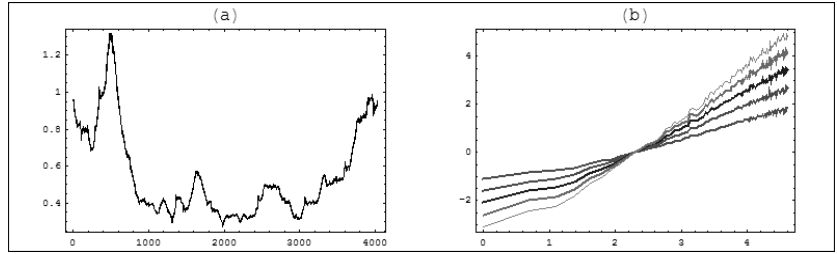}
\caption{Time series of the price (a) and plot of $\log E[\mid
x(t,\Delta t) \mid ^q]$ against $\log [\Delta t]$ with
price-fundamental feedback (b).} \label{F5}
\end{figure}

In this section we will first
 analyze the system dynamics and then we will study the variations in dynamics as some key parameters
are changed. In \emph{Figure (\ref{F2})} there are depicted the time
series of the last 500 observations of prices of S$\&$P500 and
model, returns, demands, risk attitudes and the projections of the
phase space on the planes $[D^F,\varsigma ^F]$, $[D^{TF},\varsigma
^{TF}]$, $[D^C,\varsigma ^C]$. \emph{Tables (3,4,5)} show statistics
for different parameters values. The demands of technical traders
switch between positive and negative phases, differently from the
fundamentalist demand, which instead tends to move around zero. The
presence of long phases of positive and negative demands of
technical traders, together with the dynamics for the risk aversions
may determine very large price oscillations in both directions. The
increase in the fundamental value triggers a stock price increase
due to the purchases by fundamentalists, which is reinforced by the
action of trend followers. The demand of fundamentalists has smaller
oscillations in the periods where the risk aversion is high, because
a high risk aversion induces the fundamentalists not to open large
positions if the stock is mispriced. Whereas the risk aversion of
fundamentalists follows well defined trends and is on average
positive, those of technical traders tends to oscillate around zero.
As such, technical traders switch between phases in which they are
risk averse and phases in which are risk seekers. The dynamics for
the risk attitudes may be explained in the following way: let us
assume that the price is rising and the demand of trend followers is
positive and greater than $\sqrt{V^{TF}}$. {\it Equation
(\ref{DRT})} implies that their risk aversion rises as well. The
increase in price reduces the demand of fundamentalists and
contrarians, but reinforces that of trend followers, which on the
other hand tends to fall because of the increase in their risk
aversion. Once the price falls, the demand of trend followers
approaches zero (eventually becoming negative) and, as a
consequence, their risk aversion falls. The dynamics are also the
same in the case where the cycle is triggered by fundamentalists or
contrarians. Risk attitudes may vary considerably even during phases
in which the demands are almost steady. Indeed it is sufficient that
the absolute value of the demand of investors type $i$ remains for a
long time respectively above $\sqrt{V^i}$ to get a considerable
change in risk aversion. The time derivatives of the risk attitudes
tend to reach their lower bounds, which are respectively equal to
$-\delta ^FV^F$,
 $-\delta ^{TF}V^{TF}$ and $-\delta ^CV^C$, only when the demands are
very close to zero.
\\
\\
{\bf 4.1. Effects of changing the proportion of fundamentalists and
technical traders. }In order to analyze the effect of the proportion
of fundamentalists and technical traders, we select values of
$\alpha$ ranging from 0 to 1 and with a difference of 0.1 between a
simulation and the next. If there are no fundamentalists or if their
proportion is only ten percent, the price goes to infinity, because
technical trading drives the price away from the
fundamental.\footnote{ The price goes to zero with other parameter
values. What matters here is that the price does not match the
fundamental in the long run.} If $\alpha=0.1$ the fundamentalists
are able to steer the price to the fundamental value, but prices are
subject to large oscillations induced by technical traders. Such
oscillations become larger and larger as time goes on. In fact
larger departures from the fundamental value are needed for the
fundamentalists to bring the price back close to the fundamental
value. When $\alpha=0.2$ the departure from the fundamental value
brings about long phases in which the fundamentalists go either long
or short on the asset, determining in this way an increase in their
risk aversion. This in turn implies a lower capability of offsetting
technical traders. The overall demand of the latter presents long
phases in which the demand is either positive or negative, phases in
which it changes sign quickly and phases where the demands of
contrarians and trend followers synchronize and offset each other.
During phases of synchronization the system reduces by one
dimension. When the technical demand is equal or close to zero,
fundamentalists bring the price back close to the fundamental value.
As a consequence of the fact that the total demand does not change
sign for long periods, the price tends to follow a monotonic
trajectory when it is far from the fundamental and to oscillate as
it gets close to it. Thus, the synchronization of technical traders
determines an intermittent behavior in the system with regular
monotonic phases interrupted by chaotic bursts. The time series of
fundamentalist and technical demands are depicted in {\it
\emph{Table (\ref{F3})}}.\footnote{The Lyapunov exponent is not
reported for $\alpha =0.1,0.2$ because is meaningless when the
dynamics are not bounded.} If $\alpha=0.3$ the proportion of
fundamentalist is sufficiently high as to prevent technical trading
from bringing about larger and larger departures from the
fundamental value. The oscillations have anyway larger amplitudes
than in the case where $\alpha=0.4$, and this in turn determines an
increase in the variance and a decrease in the kurtosis. If
fundamentalists account for half of the investors, the demand of
technical traders is generally lower than in the baseline case
because fundamental trading prevents strong changes in the price.
This leaves little room for a persistent phase of fundamentalist
demand and therefore fundamentalists are more likely to became risk
seekers. The higher proportion of fundamentalists determines a more
regular behavior of the system, as denoted by the decrease in
kurtosis. If the fraction of fundamentalists is equal to or greater
than sixty percent, the system no longer converges to a strange
attractor, but to a quasi-periodic attractor, as denoted by the
values of the Lyapunov exponents. If there are only fundamentalists
the attractor becomes strange again and the Lyapunov exponent rises
up to 0.53689, which would indicate a highly chaotic system. However
the rise in the Lyapunov exponent is due to the increase in the
amplitudes of the oscillations that in turn are due to the
overreaction induced by the delayed reaction of fundamentalists,
which brings price above (below) the fundamental price when the
security is originally underpriced (overpriced).
\\
\\
{\bf 4.2. Effects of changing the speed of expected price adjustment
of fundamentalists.} Increasing the speed reaction of
fundamentalists brings about a decrease in the variance because the
price tends to stay close to the fundamental. The system undergoes a
global bifurcation as the parameter $\lambda ^F$ is increased,
indeed the dynamics show a cyclical behavior after a transient
chaotic phase. This kind of transition, called attractor
destruction, is a type of crisis-induced intermittency and has been
investigated by \cite{GOY} and \cite{GORY}. However, for large
values of $\lambda^F$ the attractor becomes strange again. Because
of the presence of technical traders, which are affected by the
changes in prices triggered by the fundamentalists, it is not
possible to determine what the dynamics eventually are as the
reaction speed of the fundamentalists is further increased. For
instance, if $\lambda ^F=190$ the dynamics are periodic, but if if
$\lambda ^F=300$ the attractor is strange, with a Lyapunov exponent
of 0.240495. The projections of a limit cycle to which the system
converges when $\lambda ^F=190$ are represented in \emph{Figure
(\ref{F4})}
\\
\\
{\bf 4.3. Effects of switching between trend following and
contrarian strategies.} So far we have dealt with a model where the
proportion between trend followers and contrarians are kept
constant. If $z>0$ such proportions become path dependent. The
higher the value of $z$, the higher the fraction of trend followers
because this strategy is generally more profitable than the
contrarian one, since price grows in the long run. Simulations for
different values of $z$ show that a higher proportion of trend
followers causes greater departures from the fundamental value
triggering a reaction by all types of investors. Such dynamics bring
about an increase in the variance and skewness of returns. Skewness
tends to increase because overshooting is positive on average, since
price tends to follow an exponentially growing fundamental. Kurtosis
first tends to increase and then to decrease because the increase in
variance for high values of $z$ determines that some returns
previously in the tails of the distribution now approach the center.
\\
\\
{\bf 4.4. Effects of introducing a feedback between price and
fundamental.} We will assume now that the fundamental value is
affected by the asset price. The economic rationale is that a higher
price boosts consumption and, as a consequence, the real economy as
a whole. We assume that the dynamics of the fundamental follows the
differential equation:
\begin{equation}\label{fundamental}
\frac {\dot{F}(t)}{F(t)}=g+m \frac {P(t)}{F(t)}\,\,;\,m=0.5.
\end{equation}
The introduction of this kind of feedback induces a unit root
behavior in the price time series with scaling properties very to
those of S{\&}P500. This is apparent from \emph{Figure (\ref{F5})}
where there are reported the simulated time series and the plot of
$\log E[\mid x(t,\Delta t)\mid^q]$ against $\log [\Delta t]$ and
from the regression analysis. Indeed the $R^2$ values are
$R^2(q+1=1)=0.986382$, $R^2(q+1=1.5)=0.987099$,
$R^2(q+1=2)=0.987352$, $R^2(q+1=2.5)=0.987521$,
$R^2(q+1=3)=0.987641$.

\section{\label{sec:conclusion}Conclusion}In this paper we have outlined a continuous time deterministic model
of a financial market with heterogeneous interacting agents. The
dynamical system is able to generate some stylized facts present in
real markets, even in a purely deterministic setting: excess
kurtosis, volatility clustering and long memory. Even in the case
where fundamentalists are the only agents present in the market,
they are unable to drive the price back to the fundamental on a
steady state trajectory. Moreover, the increase in the
fundamentalist reaction speed may even increase the disorder in the
system, because the fundamentalists trigger a strong response of
technical traders. It may also be possible that, when the fraction
of fundamentalists is low, trend followers and contrarians give rise
to synchronization in the system, bringing about a dramatic change
in the dynamics. The introduction of an evolutionary switching
between technical traders leads to an increase in the volatility and
in the kurtosis, provided that the speed of switching is not too
high because otherwise the increase in the variance makes it less
likely that returns will fall in the tails of the distributions.
Further research will take into account more realistic distribution
functions for the agents, the introduction of stochastic
disturbances and a deeper investigation of the interaction between
price and fundamental.

\begin{table} \
\begin{tabular}{|c|c|c|c|c|c|c|c|}
\hline
  \textbf{} & \textbf{Mean} & \textbf{Variance} & \textbf{Skewness} & \textbf{Kurtosis} & \textbf{Jar.Bera} & \textbf{Lyap.exp.} \\
\hline
   S$\&$P500   & 0.0003597     & 0.0001026        & -0.0146       &  6.700         &  421.9          &    \\
   Model   & 0.0003617    & 0.0001100        &  -0.0293        &  6.115        &  1632          &  0.2500   \\
\hline
\end{tabular}
\caption{Statistics of S$\&$P500 and simulated time series.}
\end{table}

\begin{table} \}
\begin{tabular}{|c|c|c|c|c|c|c|}
\hline
 \textbf{$R^2$} & \textbf{$q+1=1$} & \textbf{$q+1=1.5$} & \textbf{$q+1=2$} & \textbf{$q+1=2.5$} & \textbf{$q+1=3$} \\
\hline
   S$\&$P500   & 0.9870     & 0.9854        & 0.9820       &  0.9771         &  0.9707              \\
   Model   & 0.848    & 0.8287    &  0.7980        &  0.7492        &  0.6769           \\
\hline

\end{tabular}
\caption{$R^2$ of $\log E[\mid P(t,\Delta)\mid^q]$ regressed against
$\log \Delta (t)$}
\end{table}

\begin{table} \
\begin{tabular}{|c|c|c|c|c|c|c|c|}
\hline
  \textbf{$\alpha$} & \textbf{Mean} & \textbf{Variance} & \textbf{Skewness} & \textbf{Kurtosis} & \textbf{Jar.Bera} & \textbf{Lyap.exp.} \\
\hline
   0.2   & 0.004344     & 0.00527        & 0.7748       &  3.330         &  421.9           &    \\
   0.3   & 0.00088     & 0.001154        &  0.1378        &  4.107         &  218           &  0.269   \\
   0.4   & 0.0003631     & 0.0001100       &  -0.02968        &6.115          & 1631            & 0.2500  \\
   0.5   & 0.0003317    &  0.00004472       &  0.2504        &  5.153         &  821.1           & 0.1718   \\
   0.6   & 0.0004837     & 0.0003519        & 0.02186         & 1.595        & 331.9            & 0.1118  \\
   0.7   &0.0005229      &0.0004317       & 0.01568         & 1.514         & 370.8            &  0.03375  \\
   0.8   &0.000437      & 0.0002437        & 0.01894         & 1.774         &  252.7           &  0.03621  \\
   0.9   &0.0004538      & 0.0002923        & 0.130         & 6.439         & 1999            & 0.03992  \\
   1   &0.0005047      & 0.0003806        & 0.6031         & 22.05         & 61275            &  0.536  \\
\hline
\end{tabular}
\caption{Statistics of the simulated time series as $\alpha$ varies
from 0.2 to 1.}
\end{table}

\begin{table} \
\begin{tabular}{|c|c|c|c|c|c|c|c|}
\hline
  \textbf{$\lambda ^F$} & \textbf{Mean} & \textbf{Variance} & \textbf{Skewness} & \textbf{Kurtosis} & \textbf{Jar.Bera} & \textbf{Lyap.exp.} \\
\hline
   19/15   & 0.0005586     & 0.000495        &  0.1102       &  3.876         &  137.1           &  0.2446  \\
   38/15   & 0.0004701     & 0.0003267        &  0.134        &  4.030         &  190.4           &  0.2222   \\
   57/15   & 0.0004320     & 0.0002342       &  -0.01053        &3.660          & 73.46            & 0.2639  \\
   76/15   & 0.0003842    &  0.0001536       &  0.02541        &  3.694         &  81.51           & 0.248   \\
   95/15   & 0.0003631     & 0.0001100        & -0.02968         & 6.115        & 1631            & 0.2500  \\
   114/15   &0.0003550      &0.00009703       & 0.05448         & 6.398         & 1942            &  0.2242  \\
   133/15   &0.0003584      & 0.0001020        & 0.1003         & 4.627         &  451.7           &  0.05490  \\
   152/15   &0.0003565      & 0.0001000        & 0.04832         & 4.922         & 622.6            & 0.2196  \\
   171/15   &0.0003468      & 0.00007810        & -0.155         & 1.819         & 250.6            &  0.2118  \\
   190/15   &0.00034      & 0.00007369        & 0.0004368         & 5.462      & 1019            & 0.002247  \\
   190   &0.0003355      &0.00005448         &-0.06733        &1.931          &194.9          &0.07157                \\
   300   &0.0004425      &0.0002832         &0.2366         &3.589          &96.09           &0.2404                \\
\hline
\end{tabular}
\caption{Statistics of the simulated time series as $\lambda ^F$
varies from 19/15 to 190/15 and $\lambda ^F=190\,;\,300$.}
\end{table}

\begin{table} \
\begin{tabular}{|c|c|c|c|c|c|c|c|}
\hline
  \textbf{z} & \textbf{Mean} & \textbf{Variance} & \textbf{Skewness} & \textbf{Kurtosis} & \textbf{Jar.Bera} \\
\hline
   5   & 0.000359     & 0.0001042        & 0.06756       &  6.268         &  1798            \\
   10   & 0.0003588     & 0.0001083        &  0.1103        &  5.439         &  1007          \\
   20   & 0.0003643     & 0.0001146       &  0.08794        &5.663          & 1197           \\
   30   & 0.0003838    &  0.0001539       &  0.1881        &  10.533         &  9560          \\
   40   & 0.0003900     & 0.0001465        & 0.1540         & 8.243        & 4635           \\
   60   &0.0004234      &0.0002252       & 0.3276         & 7.064         & 2848           \\
   80   &0.0004667      & 0.0003244        & 0.3391         & 7.810         &  3965            \\
\hline
\end{tabular}
\caption{Statistics of the simulated time series as $z$ varies from
5 to 80.}
\end{table}

\end{document}